\documentclass[a4paper]{jpconf}
\usepackage{graphicx}
\usepackage{color}
\usepackage{epsfig}
\usepackage{amsfonts}
\usepackage{cite}
\newcommand{\be}{\begin{equation}}
\newcommand{\ee}{\end{equation}}
\newcommand{\ba}{\begin{eqnarray}}
\newcommand{\ea}{\end{eqnarray}}
\newcommand{\nn}{\nonumber\\}
\begin{document}
\title{Strong Interactions, (De)coherence and Quarkonia}
\author{Stefano Bellucci$^1$}
\address{$^1$INFN-Laboratori Nazionali di Frascati, Via E. Fermi 40, 00044 Frascati, Italy.}
\ead{bellucci@lnf.infn.it}
\author{Vinod Chandra$^2$}
\address{$^2$Department of Theoretical Physics, Tata Institute of Fundamental Research, Homi Bhabha Road Mumbai-400005, India.}
\ead{joshi.vinod@gmail.com}
\author{Bhupendra Nath Tiwari$^1$}
\address{$^1$INFN-Laboratori Nazionali di Frascati, Via E. Fermi 40, 00044 Frascati, Italy.}
\ead{tiwari@lnf.infn.it}
\begin{abstract}
Quarkonia are the central objects to explore the non-perturbative
nature of non-abelian gauge theories. We describe the
confinement-deconfinement phases for heavy quarkonia in a hot QCD
medium and thereby the statistical nature of the inter-quark
forces. In the sense of one-loop quantum effects, we propose that
the ``quantum" nature of quark matters follows directly from the
thermodynamic consideration of Richardson potential. Thereby we
gain an understanding of the formation of hot and dense states of
quark gluon plasma matter in heavy ion collisions and the early
universe. In the case of the non-abelian theory, the consideration
of the Sudhakov form factor turns out to be an efficient tool for
soft gluons. In the limit of the Block-Nordsieck resummation, the
strong coupling obtained from the Sudhakov form factor yields the
statistical nature of hadronic bound states, e.g. kaons and Ds
particles.\\



\end{abstract}
\section{Introduction}
In this work, we study the geometric nature of the quark matter
formation. Specifically, we shall illustrate that the components
of the vacuum fluctuations define a set of local pair correlations
against the vacuum parameters, e.g. charge, mass and angular
momentum. Our consideration follows from the notion of the
thermodynamic geometry \cite{wein,rup1,rup2}. Importantly, this
framework provides a mathematical platform to exactly understand
the nature of the pair local correlations and underlying geometric
structures pertaining to the global phase transitions in
quarkonium systems. This perspective yields a well-known
understanding for the phase structures of mixtures of gases, black
holes in string theory \cite{bnt1,bnt2,rup3,aman} and in other
diverse contexts, as well.

The main purpose of the present investigation is to determine the
thermodynamic properties of the quarkonium configurations, in
general. Quantum chromodynamics (QCD), as the theory of strong
interactions, celebrates physics \cite{qcd1,qcd2,qcdn} at both the
high and low temperature domains. Thereby, our consideration plays
a crucial role in understanding the phases and stability of the
matter formation. In the soft gluon limit, viz., for a small
transverse momentum $k_{\bot}$, the behavior of the abelian theory
follows directly from the Poisson distribution and its numerical
counterparts. On the other hand, the statistical nature of the
non-abelian soft gluons is understood in terms of the Sudhakov
form factor \cite{Collins}. Let us recall that the QCD coupling
$\alpha_s(k_\bot)$ never lies near the limit $k_\bot\rightarrow
0$, and so the QCD effects are limited for the bound state thus
formed after the resummation. As mentioned in the
Ref.\cite{bullquark}, an integration over $k_\bot$ requires the
Sudhakov form factor
\begin{scriptsize} \ba h(b)= \int \frac{dk_l}{2k} \frac{dk_{\bot}}{k_{\bot}^2}
(1-\exp(-i\overrightarrow{k_{\bot}}.\overrightarrow{b})), \ea
\end{scriptsize} where $k_{\bot} \in (0, m_{P/2})$ is due to a
physical reason, i.e. that on average each soft gluon can take as
much as half of the initial center of mass energy. Towards the
determination of the index $p$, an interesting argument follows
from the work of Polyakov \cite{Polyakov} which we shall explore
further from the perspective of thermodynamic geometry in the
subsequent consideration. Before doing so, let us consider the
joint effects of the (i) confinement and (ii) rotation, and thus
make a platform to describe the thermodynamic geometry of rotating
quarkonia. Such a simplest configuration is described by Regge
trajectories with the leading order effective potential
\begin{scriptsize} \ba \label{Regge} V(r,J)=
\frac{J(J+1)}{r^2}+Cr^{2p-1}. \ea
\end{scriptsize}   In this case, the effective theory
\cite{Durand, Kuhn} is inspired from the limiting QCD strong
coupling  \begin{scriptsize} \ba \label{eq2} \alpha_s(Q^2)=
b^{-1}\frac{p}{\ln(1+p(\frac{Q^2} {\Lambda_{QCD}^2})^p)},\ea
\end{scriptsize} where $b:= (33-2N_f)/12 \pi$. In the sense of
one-loop quantum effects, we propose that the ``quantum" nature of
quark matters follows directly from the thermodynamic intrinsic
geometry of the Richardson type potentials. In the momentum space,
the net effective potential offers the right quarkonium bound
states, after taking account of the one-loop exchange terms.

Let us now focus our attention on the thermodynamic geometry of
quarkonium bound states with finitely many parameters of the
effective field theory. We consider here a framework of the
intrinsic Riemannian geometry whose covariant metric tensor is
defined as the Hessian matrix of the QCD coupling, with respect to
a finite number of arbitrary parameters carried by the soft gluons
and quarks. Such a consideration yields the space spanned by $n$
parameters of the strong QCD coupling $\alpha_s$, which, in the
present treatment, exhibits a n-dimensional intrinsic Riemannian
manifold $ M_n $. As per the definition of the thermodynamic
geometry \cite{wein,rup1,rup2,bnt1,bnt2,rup3,aman}, the components
of the covariant metric tensor are given by \be \label{eq3}
g_{ij}:=\frac{\partial^2 \alpha_s(\vec{x})}{\partial x^j \partial
x^i}, \ee where the vector $\vec{x} \in M_n $. In the strongly
coupled quarkonium effective configuration, there are only a few
physical parameters, what makes the analysis fairly simple. As
mentioned in the Ref.\cite{bullquark}, the variables of the
interest of the above quarkonium configurations are the momentum
scale parameter, $Q^2:=q$, the mass $M$, and the angular momentum
$J$, if any.
\section{Massless Quarkonia}
Let us first examine the massless non-rotating quarkonia and study
the thermodynamic stability properties as emphasized earlier, and
subsequently include the rotation.
%
%
Considering Eqn.(\ref{eq2}), one obtains the following expression
\begin{scriptsize} \ba A(q,p) := \frac{p}{b \ln(1+p(q/L)^p)}. \ea \end{scriptsize}
for the strong QCD coupling, where $L:=\Lambda_{QCD}^2$. To
compute the thermodynamic metric tensor in the space of the
parameters $\{ q, p \}$, we employ the Eqn.(\ref{eq3}), which
leads to the following expression for the components of the metric
tensor \begin{scriptsize} \ba g_{qq} &=& \frac{p^3}{bq^2}
\frac{n^Q_{11}}{r^Q_{11}}, \ \ g_{qp} = \frac{p^2}{b
q}\frac{n^Q_{12}}{r^Q_{12}}, \ \ g_{pp}=\frac{1}{b}
\frac{n^Q_{22}}{r^Q_{22}}. \ea \end{scriptsize} In order to
simplify the subsequent notations, let us define the logarithmic
factor as  \begin{scriptsize} \ba l(p):=\ln(1+p (q/L)^p). \ea
\end{scriptsize} Thus, the factors in the numerator of the local
pair correlations can be expressed as
\begin{scriptsize} \ba n^Q_{11}&:=& 2 p^2 (q/L)^{2 p}+
l(p)((q/L)^p- p(q/L)^p +(q/L)^{2 p} p), \nn n^Q_{12}&:=& 2
(q/L)^{2 p}p+ l(p)(2 (q/L)^{2p} p^2 \ln(q/L)-3 (q/L)^p -2 (q/L)^{2
p} p -(q/L)^p p \ln(q/L)), \nn n^Q_{22}&:=& 2 (q/L)^{2 p}( p +
2p^2\ln(q/L) + p^3 \ln(q/L)^2)- l(p)( (q/L)^p +(q/L)^{2 p} p \nn
&& +4 (q/L)^p p \ln(q/L)+2 (q/L)^{2 p} p^2 \ln(q/L)+(q/L)^p
p^2\ln(q/L)^2). \ea
\end{scriptsize} In the case of the massless non-rotating
quarkonia, we find that the factors $\{r^Q_{ii}| \ i=1,2,3\}$ take
the uniform value $ l(p)^3 \exp{(2 l(p))}$ in all denominators of
the local pair correlation functions. In a given QCD phase, this
happens when the parameters $\{ q, p \}$ are confined in the
domain
\begin{scriptsize} \ba\{\mathcal D:= (q,p) \in \ M_2| \
n^Q_{11}>0, \ n^Q_{22}>0 \}.\ea
\end{scriptsize} Over the above domain of $\{ q, p \}$,
the massless non-rotating quarkonia is well-behaved and locally
stable. From the definition of the thermodynamic geometry, we find
further for the generic value of the parameters that the Gaussian
fluctuations form a stable set of correlations over $\{ q, p \}$,
if the determinant of the metric tensor  \begin{scriptsize} \ba
\label{deteqQ} \Vert g \Vert = \frac{p^3}{b^2 q^2 l(p)^5 \exp{(3
l(p))}} n^Q_g \ea \end{scriptsize} remains a positive function on
the intrinsic $qp$-surface $(M_2(R),g)$. Explicitly, we obtain
that the numerator of the determinant of the metric tensor can be
expressed as  \begin{scriptsize} \ba n^Q_g(q,p):&=& 2 (q/L)^{3
p}(p+3 p^2+2 p^2 ln(q/L)+2 p^3 ln(q/L)+ p^3 ln(q/L)^2)\nn &&-
l(p)(2 (q/L)^{2 p}+ p(q/L)^{3 p}+2 p^2 (q/L)^{3 p} ln(q/L)+4 p
(q/L)^{2 p} ln(q/L)\nn &&+ p^2 (q/L)^{2 p} ln(q/L)^2 +2p^2
(q/L)^{2 p} ln(q/L) +4p^2 (q/L)^{3 p}+7 p (q/L)^{2 p}).\ea
\end{scriptsize}
In this case, we find that the scalar curvature reduces to the
following specific form  \begin{scriptsize} \ba R(q,p) = \frac{b
l(p)}{2p^2 (n^Q_g)^2} (n^{(0)Q}_R+n^{(1)Q}_Rl(p)
+n^{(2)Q}_Rl(p)^2+n^{(3)Q}_Rl(p)^3). \ea \end{scriptsize}   We
find that the factors of the numerator of the scalar curvature
take a set of interesting expressions \cite{bullquark}. In the
case when the Ricci scalar curvature $R(q,p)$ vanishes, the
underlying quarkonium system is found to be in equilibrium. Such a
state of the configuration can arise with $\{ n^{(i)Q}=0, i=
0,1,2,3 \}$, if the other factors of the scalar curvature remain
non-zero. In the other case, when the scalar curvature $R(q,p)$
diverges, the configuration goes over a transition. Such an
extreme behavior of the quarkonia is expected to happen, when
either the index $p$ or the numerator of the determinant of the
metric tensor $n^Q_g$ vanish. We further observe that the
stability of massless non-rotating quarkonia exists in certain
bands. A closer view shows that the quarkonia are decaying and
interacting particles in the Coulombic limit. For the regime of
the rising potential, the respective determinant of the metric
tensor and scalar curvature shows that the limiting rising
potential quarkonia are stable and non-interacting particles.
In the Regge trajectory model \cite{bullquark}, we find
interestingly that all possible local and global thermodynamic
stability behavior of the three parameter $\{ q, p, J \}$ massless
quarkonia remains the same up to the sign of $b_1$, as if there
were no effects of the rotation in the underlying configuration.
The fact that the efficiency of the rotation induces a mass to the
quarkonium is analyzed by considering the Bloch-Nordsieck
resummation of the angular phases.
\section{Massive Quarkonia}
Let us firstly illustrate the cases for the two parameter
configurations with either $\{q,J\}$ or $\{q,m\}$ fluctuating and
then systematically extend the consideration for the generic three
parameter quarkonia.
\subsection{Two Parameter Quarkonia}
For the of massive quarkonia, the resummed strong QCD coupling
takes the following form  \begin{scriptsize} \ba A(q,J) :=
\frac{1}{b} \frac{p (1-J(0,a \sqrt{q}))}{\ln(1+p (q/L)^p)}
\ln(\frac{\sqrt{J} +\sqrt{J-q}}{\sqrt{J}-\sqrt{J-q}}), \ea
\end{scriptsize} where $J(\nu,x)$ is the Bessel function of the
first kind of the order $\nu$. In order to describe fluctuations
in the QJ-plane, let the logarithmic factor concerning the
Bloch-Nordsieck rotation be defined as
\begin{scriptsize} \ba f(q,J):= \ln(\frac{\sqrt{J}
+\sqrt{J-q}}{\sqrt{J} -\sqrt{J-q}}).\ea \end{scriptsize} From the
Eqn.(\ref{eq3}), we find that the components of the metric tensor
are  \begin{scriptsize} \ba g_{qq}= -\frac{p( n^{(0)J}_{11}+
n^{(1)J}_{11} l(p)+ n^{(2)J}_{11} l(p)^2)}{4 b l(p)^3
\exp{(2l(p))}q^{5/2} (J-q)^{3/2}}, \ g_{qp} = \frac{p(
n^{(0)J}_{12}+ n^{(1)J}_{12} l(p))}{2 b l(p)^2 \exp{(l(p))}
q^{3/2} \sqrt{J} (J-q)^{3/2}}, \ g_{pp}=- \frac{p (2J-q) (1- J(0,a
\sqrt{q}))}{2 b l(p)J^{3/2} (J-q)^{3/2}}. \ea
\end{scriptsize}  Without any approximation, the factors in the
numerator of the pure $qq$ and $qJ$-components can be expressed as
linear combinations of the integer powers of the scaling
$(q/L)^{np}$, where $n \in Z$. As a result, we see that the
geometric nature of the parametric pair correlation functions
turns out to be remarkably interesting, viz., the domains of the  local
stability of the fluctuating quarkonia may be easily described in
terms of the parameter $q$ and $J$. Under the Gaussian
fluctuations of $\{q,J\}$, the local stability of the system
requires that (i) $qq$- fluctuations satisfy the constraint
\begin{scriptsize} \ba n^{(0)J}_{11}+ n^{(1)J}_{11}l(p)+
n^{(2)J}_{11}l(p)^2 <0 \ea \end{scriptsize} and (ii) $JJ$-
fluctuations be constrained to the following limiting values of
the Bessel function  \begin{scriptsize} \ba J(0,a \sqrt{q})&<& \
1,\ \ q>2J, \nn &>& \ 1, \ \ q<2J. \ea
\end{scriptsize}
In this case, we find that the determinant of the metric tensor
reduces to the following expression  \begin{scriptsize} \ba \Vert
g \Vert= -\frac{p^2(n^{(0)J}_{g}+ n^{(1)J}_{g}l(p)+
n^{(2)J}_{g}l(p)^2)}{8 b^2 l(p)^4 \exp{(2 l(p))} q^{5/2}J^{3/2}
(J-q)^{3/2}}, \ea \end{scriptsize} where $ n^{(1)J}_{g}=4 p^2
(q/L)^p (n^{(12)J}_{g}+ n^{(13)J}_{g}p(q/L)^p) $ and
$n^{(2)J}_{g}= n^{(20)J}_{g}+ 2 n^{(21)J}_{g} p (q/L)^p+
n^{(23)J}_{g}(q/L)^{2p} p^2$. Furthermore, it turns out that all
factors appearing in the numerator of the determinant of the
metric tensor, e.g., $\{ n^{(0)J}_{g}, n^{(12)J}_{g},
n^{(13)J}_{g}, n^{(20)J}_{g}, n^{(21)J}_{g}, n^{(23)J}_{g} \}$,
can be presented as linear  combinations of the scaling
$(q/L)^{np}$, where $n \in Z$. Combining the effects of all
fluctuations of the $\{ q,J \}$, we observe that the quarkonia are
stable for $q:= Q^2 \in (1,4)$. In general, the global stability
requires that the determinant of the metric tensor must be
positive definite, which in the present case transform as
\begin{scriptsize} \ba n^{J}_g:= n^{(0)J}_{g}+ n^{(1)J}_{g}l(p)+
n^{(2)J}_{g}l(p)^2 &<& \ 0. \ea
\end{scriptsize} It turns out that the thermodynamic curvature may
be written as the series of the charmonium logarithmic factor
$l(p)$ and Bloch-Nordsieck logarithmic factor $f(q,J)$ of rotation
as the coefficient of the expansion. Systematically, the exact
expression for the scalar curvature takes the form
\begin{scriptsize} \ba R(q,J) = \frac{b l(p)}{2p^2 (n^{J}_g)^2}
\sum_n B_n \times (l(p))^n, \ea \end{scriptsize}   where the $B_n$
in the numerator of the scalar curvature are polynomials in $p$,
whose coefficients are the functions of the Bloch-Nordsieck
logarithmic factor $f(q,J)$. In sequel, we find that the
quantitative properties of the scalar curvature and the Riemann
curvature tensor remain similar \cite{bullquark}, when we consider
the variable $Q$ as the transverse momentum with the understanding
that $k=k_{\bot}$ and the mass as the other variable of the
arbitrary two parameter real quarkonium system.
The present consideration explicates the regions of the
thermodynamic (in)stability for the massive quarkonia and the
(un)stable phases of the one loop QCD. As the non-linear effects
become stronger and stronger, it turns out that the thermodynamic
instability and correlations grow further. This motivates us to
extend our analysis to the case of more general quarkonium
configurations.
\subsection{Three Parameter Quarkonia}
Finally, let us analyze the general massive rotating quarkonia,
when all parameters, viz., the scale $q$, index $p$ and angular
momentum $J$ of the theory are allowed to fluctuate. In the
framework of the Bloch-Nordsieck resummation, the strong QCD
coupling takes the following form
\begin{scriptsize} \ba A(q, p, J)= \frac{1}{b} \frac{p(1 - J(0, a
\sqrt{q}))}{\ln(1 + p (q/L)^p)} \ln(\frac{\sqrt{J} + \sqrt{J -
q}}{\sqrt{J} + \sqrt{J - q}}). \ea \end{scriptsize}   After some
simplification, we find that the components of the metric tensor
are \begin{scriptsize} \ba g_{qq}&=& -\frac{p( n^{(0)G}_{11}+
n^{(1)G}_{11} l(p)+ n^{(2)G}_{11} l(p)^2)}{4 b l(p)^3
\exp{(2l(p))} q^{5/2}(J-q)^{3/2}}, \ g_{qp}= \frac{(n^{(0)G}_{12}+
n^{(1)G}_{12} l(p)+ n^{(2)G}_{12} l(p)^2)}{2b l(p)^3
\exp{(2l(p))}q^{3/2} (J-q)^{1/2}}, \nn g_{qJ}&=& \frac{p
(n^{(0)G}_{13}+ n^{(1)G}_{13} l(p))}{2 b l(p)^2 \exp{(l(p))}
q^{3/2} (J-q)^{3/2} J^{1/2}}, \ g_{pp}= \frac{f (-1+J(0,a
\sqrt{q})) (n^{(0)G}_{22}+ n^{(1)G}_{22} l(p))}{b l(p)^3
\exp{(2l(p))}}, \nn g_{pJ}&=& -\frac{(-1+J(0,a \sqrt{q}))
(n^{(0)G}_{23}+ l(p) n^{(1)G}_{23})}{b l(p)^2 \exp{(l(p))}
(J-q)^{1/2} J^{1/2}}, \ g_{JJ}= \frac{p (-1+J(0,a \sqrt(q))) (2
J-q)}{2b l(p) (J-q)^{3/2} J^{3/2}},\ea \end{scriptsize}   where
the coefficients $\lbrace n^{(1)G}_{11}, n^{(2)G}_{11},
n^{(1)G}_{12}, n^{(2)G}_{12} \rbrace $ appearing in the components
of the metric tensor are shown to factorize as before. In the
powers of $p$, the exact factorizations are given as follows:\\
(i) the $qq$-component \begin{scriptsize} \ba n^{(1)G}_{11}&=& 4
p^2 (q/L)^p n^{(12)G}_{11} +4 p^3 (q/L)^{2p} n^{(12)G}_{11}, \nn
n^{(2)G}_{11}&=& n^{(20)G}_{11}+ 2p(q/L)^p n^{(21)G}_{11}
+p^2(q/L)^{2p},\ea \end{scriptsize} (ii) the $qp$-component
\begin{scriptsize} \ba n^{(1)G}_{12}&=& p(q/L)^p
n^{(11)G}_{12}+p^2(q/L)^{2p} n^{(12)G}_{12},\nn n^{(2)G}_{12}&=&
n^{(20)G}_{12}+ 2p(q/L)^p n^{(21)G}_{11} +p^2(q/L)^{2p}
n^{(22)G}_{12}.\ea \end{scriptsize} Interestingly, the factors
appearing in the numerator of the $qq$, $qp$ and $qJ$-components
can be cascaded as a linear combination in the integer powers of
the scaling $(q/L)^{pn}$, where $n \in Z$. Similarly, the
corresponding factors of numerator of the $pp$-components take the
following expressions
\begin{scriptsize} \ba n^{(0)G}_{22}&=& -2 p (q/L)^{2 p}(1+2p
\ln(q/L)+p^2 \ln(q/L)^2), \nn n^{(1)G}_{22}&=& (q/L)^p(2+4 p
\ln(q/L) + p^2 \ln(q/L)^2)+p(q/L)^{2 p}(1 +2 p \ln(q/L)).\ea
\end{scriptsize} Finally, the factors in the numerator of the
$pJ$-components are  \begin{scriptsize} \ba n^{(0)G}_{23}&=& -p
(q/L)^p -p^2 \ln(q/L) (q/L)^p, \nn n^{(1)G}_{23}&=& 1 + p (q/L)^p.
\ea \end{scriptsize}   The local stability of the configuration
requires a set of constraints on the domains of the
parameters. Specifically, for the same sign of $\{p, b\}$, we find
that the local stability of the fluctuating quarkonia enforces the
following simultaneous requirements
\begin{enumerate}
\item the $qq$- fluctuations satisfy  \begin{scriptsize} \ba n^{(0)G}_{11}+
n^{(1)G}_{11} l(p)+ n^{(2)G}_{11} l(p)^2 <0, \ea \end{scriptsize}
\item the $JJ$- fluctuations remain within the limiting values of
the Bessel function  \begin{scriptsize} \ba J(0,a \sqrt{q})&>& \
1,\ \ 2J<q, \nn &<& \ 1, \ \ 2J>q, \ea \end{scriptsize}   \item
the $pp$- fluctuations satisfy
\begin{scriptsize} \ba n^{(0)G}_{22}+ n^{(1)G}_{22} l(p) &>& \ 0, \ \ J(0,a \sqrt{q})
> 1, \nn &<& \ 0, \ \ J(0,a \sqrt{q}) > 1 \ea \end{scriptsize}
\end{enumerate} for the same sign of $\{f, b\}$.
The silent feature of the $JJ$-fluctuations is that we find the
two distinct local behaviors for $J>0$ and $J<0$. Subsequently, we
find that the $qp$-surface is stable, if there exits a positive
surface minor  \begin{scriptsize} \ba \label{minor} p_S^G :=
-\frac{(n^{(0)G}_{S}+ n^{(1)G}_{g}l(p)+ n^{(2)G}_{S}l(p)^2+
n^{(3)G}_{S}l(p)^3)}{4b^2 l(p)^5 \exp{(3l(p))} q^{5/2}
(J-q)^{3/2}}. \ea \end{scriptsize}   It is worth mentioning that
$\{n^{(0)G}_{S}, n^{(1)G}_{S}, n^{(2)G}_{S}, n^{(3)G}_{S}\}$ have
the following factorizations
\begin{scriptsize} \ba n^{(0)G}_{S}&=& (q/L)^{3p} n^{(03)G}_{S},
\nn n^{(1)G}_{S}&=& (q/L)^{2p} n^{(12)G}_{S} + (q/L)^{3p}
n^{(13)G}_{S}, \nn n^{(2)G}_{S}&=& (q/L)^{p} n^{(21)G}_{S}
+(q/L)^{2p} n^{(22)G}_{S}+ (q/L)^{3p} n^{(23)G}_{S}, \nn
n^{(3)G}_{S}&=& n^{(30)G}_{S} +(q/L)^{p} n^{(31)G}_{S} +(q/L)^{2p}
n^{(32)G}_{S}+ (q/L)^{3p} n^{(33)G}_{S}.\ea \end{scriptsize}   In
this identification, the factors appearing in the powers of
$(q/L)^{p}$ have the following structures
\begin{scriptsize} \ba n^{(03)G}_{S}&=& 8 f^2 n^{(032)G}_{S}, \nn
n^{(1i)G}_{S}&=& n^{(1i0)G}_{S}+ n^{(1i1)G}_{S}f+
n^{(122)G}_{S}f^2, i=2,3, \nn n^{(2i)G}_{S}&=& n^{(2i0)G}_{S}+
n^{(2i1)G}_{S}f+ n^{(2i2)G}_{S}f^2, i= 1,2,3, \nn n^{(3i)G}_{S}&=&
n^{(3i0)G}_{S}+ n^{(3i1)G}_{S}f+ n^{(3i2)G}_{S}f^2, i=0,1,2,3. \ea
\end{scriptsize} Further, the factors $032$,
$12i$, $13i$, $21i$, $22i$ and $23i$ take the following explicit
expressions
\begin{scriptsize} \ba n^{(032)G}_{S}&=& p^4 n^{(0321)G}_{S}+ p^5
n^{(0322)G}_{S}+ p^6 n^{(0323)G}_{S}, \nn n^{(12i)G}_{S}&=& p^2
n^{(12i2)G}_{S}+ 2 p^3 n^{(21i3)G}_{S}+ p^4 n^{(12i4)G}_{S}, i= 0,
1 \nn n^{(122)G}_{S}&=& p^2 n^{(2122)G}_{S}+ 2 p^3
n^{(2123)G}_{S}+ p^4 n^{(2124)G}_{S}+ 4 p^5 n^{(2125)G}_{S}, \nn
n^{(13i)G}_{S}&=& 4 p^3 n^{(13i3)G}_{S}+ 8 p^4 n^{(13i4)G}_{S}+ 4
p^5 n^{(13i5)G}_{S}, i= 1,2, \nn n^{(21i)G}_{S}&=& 2 p
n^{(21i1)G}_{S}+ 2p^2 n^{(21i2)G}_{S}+ p^3n^{(21i3)G}_{S}, i= 1,
2, \nn n^{(22i)G}_{S}&=& p^2 n^{(22i1)G}_{S} + 2 p^3
n^{(22i2)G}_{S} +p^4 n^{(22i3)G}_{S}, i= 1,2, \nn
n^{(23i)G}_{S}&=& p^3 n^{(23i3)G}_{S} + 2p^4 n^{(23i4)G}_{S}, i=
1, 2. \ea \end{scriptsize} Interestingly, the sub-factors of the
factors $032$, $120$, $121$, $122$, $130$, $131$, $211$, $212$,
$220$, $221$, $222$, $230$, $231$ and $232$ can be expressed as
the linear combinations of the integral powers of the scaling
$\ln(q/L)$ and the integral powers of the zeroth order Bessel
function. As per the expectation, we observe that the
$3ij$-factors that appear are relatively simpler than their
foregoing counterparts. In fact, we find that the $30$-factors can
be seen as a two variable polynomial in $J,q$. For the general
quarkonia with $\{q,p,J\}$ fluctuating, the stability of the
$qp$-surface requires that the principle minor $p_S^G $ remains
positive on $(M_3,g)$. Correspondingly, this leads to the
constraint that the $\{q,p,J\}$ satisfy
\begin{scriptsize} \ba n^{(0)G}_{S}+ n^{(1)G}_{g}l(p)+
n^{(2)G}_{S}l(p)^2+ n^{(3)G}_{S}l(p)^3 &<& 0. \ea \end{scriptsize}
As per the generic fluctuations of the strongly coupled massive
rotating quarkonia, we obtain the following general expressions
for the determinant of the metric tensor  \begin{scriptsize} \ba
\nn \Vert g \Vert &=& -\frac{p f (-1+ J(0,a \sqrt{q}))}{8 b^3
l(p)^6 \exp{(3 l(p))} q^{5/2} J^{3/2} (J-q)^{3/2}} (n^{(0)G}_{g}+
n^{(1)G}_{g}l(p)+ n^{(2)G}_{g}l(p)^2+ n^{(3)G}_{g}l(p)^3). \ea
\end{scriptsize} As mentioned before, we find that the
$\{n^{(0)G}_{g}, n^{(1)G}_{g}, n^{(2)G}_{g}, n^{(3)G}_{g}
\}$-terms factorize as follows
\begin{scriptsize} \ba n^{(0)G}_{g}&=& (q/L)^{3p} (8p^4
n^{(04)G}_{g}+ 8p^5 n^{(05)G}_{g}+ 8p^6 n^{(06)G}_{g}), \nn
n^{(1)G}_{g}&=& (q/L)^{2p} n^{(12)G}_{g}+ (q/L)^{3p}
n^{(13)G}_{g}, \nn n^{(2)G}_{g}&=& (q/L)^p n^{(21)G}_{g}
+((q/L)^{2p} n^{(22)G}_{g}+ (q/L)^{3p} n^{(23)G}_{g}, \nn
n^{(3)G}_{g}&=& n^{(30)G}_{g} +(q/L)^p n^{(31)G}_{g} +(q/L)^{2p}
n^{(32)G}_{g}+ (q/L)^{3p} n^{(33)G}_{g}, \ea \end{scriptsize}
After a direct simplification, we obtain the sub-factorizations
\begin{scriptsize} \ba n^{(0i)G}_{g}&=& n^{(0i0)G}_{g}+ f
n^{(0i1)G}_{g}, i=4,5,6, \nn n^{(1i)G}_{g}&=& n^{(1i0)G}_{g}+ f
n^{(1i1)G}_{g}, i=2,3, \nn n^{(2i)G}_{g}&=& n^{(2i0)G}_{g}+ f
n^{(2i1)G}_{g}, i=1, 2, 3, \nn n^{(3i)G}_{g}&=& n^{(3i0)G}_{g}+ f
n^{(3i1)G}_{g}, i=0, 1, 2, 3. \ea
\end{scriptsize} As per our computation, we observe that the
terms appearing in the various powers of $l(p)$ have a
sub-factorization in the index $p$. Specifically, we find that the
factors of $12$, $13$, $21$, $22$ and $23$-components are given by
\begin{scriptsize} \ba n^{(120)G}_{g}&=& n^{(1202)G}_{g} p^2+
n^{(1203)G}_{g} p^3+ n^{(1204)G}_{g} p^4, \nn n^{(121)G}_{g}&=&
n^{(1212)G}_{g} p^2 +n^{(1213)G}_{g} p^3 +n^{(1214)G}_{g} p^4
+n^{(1215)G}_{g} p^5, \nn n^{(130)G}_{g}&=& n^{(1303)G}_{g} p^3+
n^{(1304)G}_{g} p^4+ n^{(1305)G}_{g} p^5, \nn n^{(131)G}_{g}&=&
n^{(1313)G}_{g} p^3+ n^{(1314)G}_{g} p^4+ n^{(1315)G}_{g} p^5+
n^{(1316)G}_{g} p^6, \nn n^{(21i)G}_{g}&=& n^{(21i1)G}_{g} p+
n^{(21i2)G}_{g} p^2+ n^{(21i3)G}_{g} p^3, i= 0, 1, \nn
n^{(22i)G}_{g}&=& n^{(22i2)G}_{g} p^2+  n^{(22i3)G}_{g} p^3+
n^{(22i4)G}_{g} p^4, i=0,1, \nn n^{(23i)G}_{g}&=& n^{(23i3)G}_{g}
p^3+ n^{(23i4)G}_{g} p^4, i= 0, 1. \ea \end{scriptsize}   As in
the case of the surface minor, we observe that all the individual
sub-factorizations, (e.g. $040$, $041$, $050$, $051$, $060$,
$061$, $120i$, $121j$, $130i$, $130j$, $2303$, $2304$, $2313$,
$2314$), can be expressed as the linear combination over the
integral powers of the zeroth order Bessel function and the
scaling $\ln{(q/L)^n}$. In the case of the $l(p)^3$ terms, we find
that the sub-factors pertaining to $30$, $31$ and $32$-factors
have no dependence on  $\ln{(q/L)^n}$ and thus they are
expressible as polynomial expressions in the Bessel function only.
From this observation, we predict that the regions of the
thermodynamic stability are present for $q,J \in (1,4)$. Globally,
the stability of $(M_3,g)$ constrains the principle minors
$\{g_{ii}, p_S^G, \Vert g \Vert \}$ to remain positive.
Specifically, for the same sign of $\{ b, p, f\}$, the volume
stability of the $(M_3,g)$ imposes the following constraint
\begin{scriptsize} \ba \sum_{i=0}^3 n^{(i)G}_{g}l(p)^i &>& \ 0, \
J(0,a \sqrt{q}) < 1, \nn &<& \ 0, \ J(0,a \sqrt{q}) < 1. \ea
\end{scriptsize} Importantly, it is worth mentioning that both the
limiting configurations with $J=q$ and $J(0,a \sqrt{q})= 1$ are
abided from the thermodynamic stability constraints. To summarize
the phases of generic quarkonia, the exact formula for the scalar
curvature may analogously be deduced, as the one we have offered
for the fluctuations in the $qJ$-plane. In this sense, we find
that the summation over $l(p)$ naturally arises with the $B_n$ as
the polynomials in $p$, whose coefficients can be expressed as the
functions of the Bloch-Nordsieck logarithmic factor $f(q,J)$.
Based on the analysis of the present paper, the thermodynamically
stable index of the Bloch-Nordsieck rotating massive quarkonia is
constrained by the following set  \begin{scriptsize} \ba \mathcal
P_{m \ne 0}:= \{p \ | \ n^{G}_{ii}> 0,\ n^{G}_{S}> 0,\ n^{G}_{g}>
0 \}. \ea \end{scriptsize}  As the gluons become softer and
softer, we find, in the limit of Bloch-Nordsieck resummation, that
the underlying Sudhakov form factor offers all possible
thermodynamically stable phases of the strongly coupled quarkonia.
As examined for the $QJ$-plane, we would like to explicitly
understand properties of the set $\mathcal P_{m \ne 0}$ and the
associated Ricci scalar curvature. Up to a phase of QCD, the
global properties of the three parameter quarkonia remain the same
as we have exactly indicated for the two parameter quarkonia. As
mentioned in the Ref.\cite{bullquark}, we have computed the
intrinsic geometric properties of the Bessel function of the first
kind convoluted with two logarithmic functions of the respective
weights $(0,2p)$. Further analysis of the geometric features of
these exploitations is left for the future.
\section{Conclusion and Outlook}
We have examined the role of the thermodynamic intrinsic geometry
for a class of  quarkonium configurations. We have offered a
geometric perspective to the confinement- deconfinement phase of
(heavy) quarkonia in a hot QCD medium and thereby described the
statistical nature of the inter-quark forces. Specifically, the
intrinsic geometric analysis provides a set of physical
indications encoded in the geometric quantities, e.g., the scalar
curvature and possible geometrically non-trivial invariants offer
the global correlation properties of an ensemble or subensemble of
equilibrium configurations. In the sense of statistical mechanics,
our analysis involving a Gaussian distribution of the particles
ensures the thermodynamical properties of the underlying quarkonia
in the late time limit. From the perspective of one-loop quantum
effects, the nature of quark matter is shown to follow directly
from the thermodynamic consideration of the Richardson potential.

Our study of the quarkonia could further be explored towards other
configurations concerning the non-perturbative and non-abelian
nature of the gauge theories. Such a consideration provides a
unified description, encompassing all the regimes of QCD at finite
temperature, i.e. the Coulombic, the linear rising and the Regge
rotating regimes, for both massless and massive quarkonia.
Phenomenologically, our results can be thus be used to investigate
the statistical nature of soft gluons and the associated
phenomenon at the LHC.
\subsection*{Acknowledgments}

The work of S.B. has been supported in part by the European
Research Council grant n.~226455, \textit{``SUPERSYMMETRY, QUANTUM
GRAVITY AND GAUGE FIELDS (SUPERFIELDS)"}.

B.N.T. would like to thank organizers of the \textit{``DISCRETE
2010, Rome, Italy"} for providing stimulating environment towards
this work. B.N.T. further acknowledges postdoctoral research
fellowship of the \textit{``INFN-Laboratori Nazionali di Frascati,
Roma, Italy''}.

\end{document}